# Atomic-Scale Roughness of Freestanding Oxide Membranes Revealed by Electron Ptychography

Huaicheng Yuan[1,2], Yu-Chen Liu[3], Li-Shu Wang[3], Zehao Dong[4], Jan-Chi Yang[3,5], Zhen Chen[1,2]*

**Affiliations**

[1]Beijing National Laboratory for Condensed Matter Physics, Institute of Physics, Chinese Academy of Sciences, Beijing 100190, P. R. China

[2]School of Physical Sciences, University of Chinese Academy of Sciences, Beijing 100049, P. R. China

[3]Department of Physics, National Cheng Kung University, Tainan 70101, Taiwan

[4]State Key Laboratory of Low Dimensional Quantum Physics, Department of Physics, Tsinghua University, Beijing 100084, P. R. China

[5]Center for Quantum Frontiers of Research & Technology (QFort), National Cheng Kung University, Tainan 70101, Taiwan

*Corresponding authors. Email: zhen.chen@iphy.ac.cn

## Abstract

Freestanding oxide films offer significant potential for integrating exotic quantum functionalities with semiconductor technologies. However, their performance is critically limited by surface roughness and interfacial imperfection caused by dangling bonds, which disrupt coherent interactions and suppress quantum phenomena at heterointerfaces. To address the challenge of structural characterization of surfaces and interfaces, we develop a metrological approach achieving atomic-scale precision in mapping the topography of both free surfaces and buried interfaces within ultrathin oxide heterostructures leveraging three-dimensional structures reconstructed from multislice electron ptychography. This method also allows for counting the number of atoms, even including light elements such as oxygen, along the electron trajectory in electron microscopy, leading to the identification of surface termination in oxide films. The planar-view of measurement geometry, allowing for large field-of-view imaging, provides remarkably rich information and high statistics about the atomic-scale structural inhomogeneities in freestanding membranes. This quantitative analysis provides unprecedented capabilities for correlating structural imperfection with quantum device performance, offering critical insights for engineering robust heterointerfaces in next-generation oxide electronics.

## INTRODUCTION

Freestanding epitaxial oxide films, released from rigid substrates, retain the rich functionalities and strong electronic correlations of complex oxides, and can even host emergent phenomena, while enabling versatile integration into diverse device architectures(*1-6*). These membranes unlock unprecedented opportunities for ferroelectric memory, piezoelectric sensing and energy harvesting, and dielectric energy storage, driving innovations across information and energy technologies (*4, 7, 8*). Examples include lead titanate/strontium titanate bilayers on silicon exhibiting room-temperature skyrmion-like polar nanodomains(*9*), and freestanding $BaTiO_3$ membranes demonstrating promise for nonvolatile ferroelectric domain wall memory(*10*). However, intrinsic performance critically depends on surface and interfacial properties: Surface dangling bonds and local curvature can perturb lattice parameters, charge distributions, and orbital configurations. These perturbations can destroy the expected functional properties such as ferromagnetism, ferroelectricity, or superconductivity(*11-16*), hindering dense, high-quality integration of these membranes into functional stacks, akin to van der Waals heterostructures(*17*). Consequently, precise and universally applicable atomic-scale characterization of surface and interface topography is essential to elucidate structure-function relationships in freestanding oxide systems.

Cross-sectional scanning transmission electron microscopy (STEM) is widely used to characterize surface structures or interface separations in twisted oxides systems. However, its projection geometry superimposes all structural features along the beam path, obscuring local surface/interface topography. Planar-view STEM image provides a wider field of view and preserves intrinsic device structures but shares the same fundamental limitation: its projection nature prevents direct access to three-dimensional (3D) surface morphology and local nanoscale gaps within heterostructures—especially in defective or non-uniformly curved regions. While techniques like exit-wave reconstruction or optical sectioning STEM can extract surface and internal structures, they often fail under strong electron multiple scattering conditions (*18, 19*). Alternative methods such as atomic electron tomography, which reconstructs three-dimensional atomic structures from tilt-series images, face limitations including experimental complexity, resolution constraints, and pronounced sample radiation damage risk(*20-22*).

In contrast, multislice electron ptychography (MEP) reconstructs three-dimensional structures from four-dimensional (4D) STEM datasets, providing direct volumetric sensitivity to both surface atoms and buried interfaces (*23-25*). Here, we introduce a robust methodology that extracts atomic-scale surface and interfacial topography from MEP-reconstructed 3D phase images based on the electron-beam propagation model. This approach reliably determines 3D atomic positions in the presence of surface roughness. Crucially, surface termination layers and number of atoms along individual columns are consistently resolved through either analysis of terminal depth positions or projected phase magnitudes. Compared with the previous atomic counting methods based on high-angle annular dark-field (HAADF) images (*26-29*), this approach provides higher precision in atom counting and enhanced sensitivity to light elements such as oxygen. Validation across model systems—including single crystalline $SrTiO_3$, freestanding $SrRuO_3$ and twisted bilayer $SrTiO_3$ films—confirms the method's accuracy. By leveraging fast 4D-STEM acquisition, this technique enables high-precision nanoscale metrology over micron-scale fields of view, establishing a universal approach for atomic-scale surface and interface characterization in ultrathin functional materials and devices.

## RESULTS

### Principle and reconstruction process

In thin laminar TEM samples, high-energy electron wave traversing the specimen from vacuum accumulate the phase, $\varphi_s = \sigma V_s$, proportional to the electrostatic potential $V_s$ (where $\sigma$ is the electron voltage dependent constant). The 4D-STEM dataset captures position-dependent Fourier spectra of this exit wave, encoding information from all atoms–including surface atoms–along the beam path. Multislice electron ptychography solves the inverse scattering problem to reconstruct the 3D crystal structure by retrieving the phase change $\varphi_s$ slice-by-slice. The potential $V_s$ approaches elemental-specific finite values within the sample, while it decays to zero in vacuum (Fig. 1A). Consequently, the local 3D potential (and its MEP-derived phase) around the sample surfaces is accurately modeled as a step function convolved with a Gaussian kernel. This convolution accounts for finite depth-resolution broadening, yielding an error function (Fig. 1B). Figure 1C illustrates the depth-dependent phase profile across an atomic column of the 3D reconstructed phase using simulated data from $SrTiO_3$ model (Methods). Fitting error functions to the upper and lower surfaces yields the depth termination of the atomic column; their difference defines the local sample thickness (Methods). If the lattice parameter along the projection-axis is known, the number of atoms can be counted in each individual column. The surface termination, surface roughness and local curvature can be further determined from the spatial variations of the fitted atomic depth positions.

We validated this approach via simulations of $SrTiO_3$ with $TiO_2$ surface termination (20 TiO, 19 Sr atoms per column; Methods). The reconstructed upper and lower surface topographies yield average depths of 4.01 ± 0.04 nm, and 11.78 ± 0.03 nm, respectively. The reconstructed thickness is 7.77±0.05 nm, matches the real sample thickness of 7.81 nm very well. Atom counts derived from thickness/lattice parameter (3.905 Å) closely match the model (Fig. 1D), with minor outliers of only ±1 atom. Counts from total phase shift per unit cell also show excellent agreement, confirming $TiO_2$ termination (Fig. 1E). Consistency between both methods validates the determination of surface termination. Notably, total phase is susceptible to reconstruction artefacts and point defects (e.g., vacancies or substitutions); thus depth-derived positions were prioritized for surface termination determination. The depth difference between the TiO atoms and the average depth of their nearest-neighbor Sr atoms, $d_{\mathrm{TiO}} - d_{\mathrm{Sr}}$, further confirmed the $TiO_2$ surface termination: negative at the upper surface and positive at the lower surface for $TiO_2$ terminated surfaces (Fig.1F). Notably, reconstruction uncertainties near the edge of the reconstruction region, and potential defects have been carefully considered during the analyses (Methods).

Experimental verification used a focused ion beam (FIB)-prepared $SrTiO_3$ wedge sample exhibiting inherent surface unevenness. MEP-reconstructed slices revealed contrast variation near both surfaces (Fig. 2, A and B; all slices depicted in Movie S1). A phase profile along the depth direction exemplifies how the phase varies in three-dimensional space (Fig. 2C). The thickness map (Fig. 2D) confirmed the wedge sample shape. Sr and TiO columns were distinguished by using the phase magnitude of the speak positions (mean TiO phase statistically larger than Sr; Fig. S1). The depth different, $d_{\mathrm{TiO}} - d_{\mathrm{Sr}}$, is negative (upper surface) and positive (lower surface), confirming $TiO_2$ termination for both surfaces (Fig. 2E). Atom counting results showed one excess Ti per TiO column relative to neighboring Sr atoms (Fig. 2F), further supporting $TiO_2$ termination–consistent with previous reports for as-treated $SrTiO_3$ (001) surface, as FIB milling is preferentially removes Sr due to its weaker bonding versus Ti-O(*30*). Thickness from depth difference agrees with the summed-phase analysis (Fig. S2). Due to the presence of phase errors, the obtained results may contain up to ~5% deviation (depends on the convergence behavior of MEP reconstruction). Surface roughness of both surfaces was ~0.4 nm (1 unit cell) via the standard deviation of the surface profiles (Fig. S3). Oxygen column atom counts match TiO atom counts (Fig. S4), validating $TiO_2$ termination.

## Surface topography of freestanding oxides

We further applied the analysis to freestanding thin $SrRuO_3$ films, a correlated ferromagnetic metal that serves as a prototypical oxide electrode and a platform for emergent topological spin phenomena such as the Hall effect and skyrmions (*12, 31*). Fig. 3A shows the summed phase from a representative region A, with Fig. 3B displaying an enlarged view (all slices in Movie S2). The reconstructed upper and lower surface topographies (Figs. 3C and 3D) yield average depths of 5.28±0.41 nm, and 12.92±0.40 nm, respectively. This measured uncertainty (0.40 nm) significantly exceeds the method's intrinsic error of 0.06 nm (estimated from simulations shown in Fig. S5), confirming surface roughness as the dominant source. At ~1 unit cell variation, this roughness level represents a typical lower limit achievable for well-controlled grown freestanding oxide films(*32*).

The middle plane between the upper and lower surfaces (Fig. 3E) has a mean depth of 9.10±0.26 nm, while the thickness map (Fig. 3F) gives an average thickness of 7.64±0.62 nm. Stripe-like features in Fig. 3E show localized deviations of ~±1 nm from the mean plane without correlated thickness variations in Fig. 3F, indicating bending likely introduced during sample transfer. Figure 3G shows that $d_{\mathrm{Sr}} - d_{\mathrm{RuO}}$ is negative (upper surface) and positive (lower surface), the difference between the number of Sr atoms and the average number of their nearest-neighbor RuO atoms is ~1.3 (Fig. 3H), indicating the more stable SrO termination(*33, 34*). Also, we calculated the thickness and atom counts by summed phase, yielding consistent results (Fig. S6). Similar analyses were performed on sample region B (fig. S7, Movie S3), a 90°-twin of region A. This region exhibits comparable surface roughness (~ 1 unit cell) and identical SrO termination. Notably, we simultaneously resolved both oxygen positions and $RuO_6$ octahedra rotation angles (~5.83°)– critical parameters governing emergent phenomena in this material (*31, 35, 36*).

**Interfacial spacing of twisted bilayer $SrTiO_3$**

We also determine the interface spacing of a twisted bilayer $SrTiO_3$ sample with angstrom-level precision beyond the capabilities of previous techniques. The suspended bilayer stacks with a relative angle of 9.45° (Fig. S8) and each layer is a few nanometers thick. Conventional HAADF images show as Moiré patterns but hardly distinguish separate layers or precise interface due to the projection effect (Fig. S8). MEP can reconstruct the 3D structure slice by slice and separate the upper and lower layers. Three selected slices from a cropped region are shown in Fig. 4A-C with all slices depicted in Movie S4, and the whole reconstruction from a large field of view of 20 × 80 $nm^2$ is shown in Fig. S9. Notably, a few slices near the interface still have intermixing effects and form Moiré patterns (Fig. 4B) due to the limited depth resolution, ~ 3 nm. The inhomogeneity at a larger scale induced by surface roughness or film curvature can be seen in the large field of view image (Fig. S10). The nonuniform interfacial contact and interfacial gap variation can be indicated from the unevenness of the Moiré patterns, which will be discussed in more details below.

Since the bilayer structure can be separated from MEP, the surface topography for the upper and lower layers can be independently determined. A key challenge arises from the intermixing effect near the interface, which complicates the precise localization of the two embedded surfaces in the interface. However, the frequent misalignment between atomic columns in the first layer and in the second layer allows for the independent fitting for the atomic columns within each layer, enabling the reconstruction of the interfacial topography. Practically, when lateral atomic positions from adjacent layers, such as the intermediate regions between Moiré cores, are in close proximity, the intermixing effect can make the atomic columns elongated. Fitting the first layer might yield the maximum gradient corresponding to the lower surface of the second layer, rather than the intended surface of the first layer. Such deviations violate the assumptions of the hypothetical model, rendering those specific fits unreliable and necessitating their manual exclusion.

To enhance measurement reliability, we propose an iterative refinement method. Initial fits from unambiguously identified atomic columns are used to predict the expected positions of the upper and lower surfaces within a layer. This prediction allows for the systematic exclusion of outlier fits that clearly deviate from expectations. For example, depth of lower surface for the first layer aligns with the second layer's positions. This manual exclusion process will leave some missing data points. But for the continuous membranes without abrupt structural changes, linear interpolation using reliable points can be used to fill the missing points and construct accurate three-dimensional distribution maps of both the upper and lower surface topographies. Implementing this iterative fitting approach significantly improves the accuracy and reliability of surface topography reconstruction in multilayer materials.

Figure 4D-E illustrate the three-dimensional topographies of the first and second layers of twisted bilayer $SrTiO_3$, revealing notable surface roughness. The first layer's upper surface has an average height of 5.44±0.22 nm, and its lower surface averages 10.45±0.38 nm (Fig. 4F). The second layer's upper surface averages 11.58±0.33 nm (Fig. 4G), with its lower surface at 15.95±0.36 nm. The interlayer spacing is depicted in Fig. 4H, with a mean interface spacing of 1.31±0.39 nm, after resampling these surfaces. For Fig. 4F–H, in the upright triangle, atoms in L1 are positioned slightly above those in U2, indicating a small interfacial spacing. In contrast, within the inverted triangle region, atoms in L1 are located noticeably higher relative to U2, corresponding to a larger interfacial spacing. Since interface-induced novel emergence properties often require close contact with strong interaction(*37*), this gap between the two layers, of about three unit cells, could dramatically degrade the interact and hardly achieve the expected functionality of stacked bilayers (*16*). From the 2D map (Fig. 4H) and the statistical analysis (Fig. 4I) of the gap, small fraction of close-contact regions (~2%) are also present. This demonstrates that the technique can be used to diagnose the interfacial topology at the atomic scale with remarkable statistics, providing crucial structural insights especially for complex multilayer systems or devices (*38-42*), which cannot be realized from other projection-based STEM imaging techniques.

## DISCUSSION

We have demonstrated that quantitative analysis of MEP-reconstructed 3D phase permits atomic-scale determination of surface and interface structures in oxide membranes. This methodology achieves two long-pursued goals in electron microscopy: Precise atom counting including light oxygen atoms and atomic-scale surface morphology determination(*19, 27*). Leveraging planar-view geometry with conventional illumination enables robust statistics for nanoscale structural variations across diverse samples–particularly complex multilayer devices. This method also provides a useful structural diagnosed tool for the next-generation atomic-scale manufactured devices (*43*). Recent advances in depth resolution, whether through new algorithms(*44*) or experimental designs(*45*), promise further reduction of measurement uncertainties and enhancement of topography precision, extending to subtle 3D atomic displacements such as topological ferroelectric textures(*46*).

## Materials and Methods

### Sample growth and preparation for experiment

Bulk $SrTiO_3$ specimen was prepared from single crystal $SrTiO_3$ using a $Ga^+$ ion beam via the conventional lift-out technique on a focused ion beam system (FIB, FEI Strata 400). To minimize ion-induced damage during thinning, the beam energy was progressively decreased from 30 keV to 2 keV. The thinnest part of the sample is less than 20 nm with a wedge shape.

Freestanding membranes were released from films grown using pulsed laser deposition (PLD) with a KrF excimer laser (248 nm). The $SrRuO_3$ films were initially grown on (La,Sr)$MnO_3$ (LSMO) buffer (001)-oriented $SrTiO_3$ substrates, deposition conditions included an oxygen ambient of 100 mTorr, a temperature of 700 °C, a laser energy of 250 mJ per pulse, and a repetition rate of 10 Hz. After growth, the samples were cooled in an oxygen pressure of approximately 700 Torr to minimize oxygen vacancies. To release the $SrRuO_3$ film, the LSMO layer serving as a sacrificial layer was etched away in hydrochloric acid, producing a detached $SrRuO_3$ membrane. The $SrTiO_3$ films were initially grown on LSMO buffer (001)-oriented $SrTiO_3$ substrates under deposition conditions identical to those used for $SrRuO_3$. To release the $SrTiO_3$ film, the LSMO layer serving as a sacrificial layer was etched away in hydrochloric acid, producing the freestanding $SrTiO_3$ membranes. The final $SrTiO_3$ twisted bilayer membrane was constructed by stacking two freestanding $SrTiO_3$ films with a deliberately introduced twist angle directly on a copper TEM grid, facilitating subsequent STEM characterization. The interface of the two layers is clean as shown in the Supplement Video 4.

### 4D-STEM experiments

The experiments involving FIB–fabricated $SrTiO_3$ and freestanding $SrRuO_3$ were performed using a double aberration-corrected scanning transmission electron microscope (STEM, Thermo Fisher Scientific, Titan Themis) operated at an accelerating voltage of 300 kV. A high-dynamic-range electron microscope pixel array detector (EMPAD) was employed to acquire a four-dimensional STEM dataset with a data volume of 124 × 124 × 200 × 200 for $SrRuO_3$ and 124 × 124 × 128 × 128 for $SrTiO_3$. The scan step size was 0.474 Å for $SrRuO_3$ and 0.363 Å for $SrTiO_3$, and the convergence semi-angle was set to 25.2 mrad for $SrRuO_3$ and 24 mrad for $SrTiO_3$. The electron probe current was maintained at 33 pA, with a dwell time of 1.86 ms per diffraction pattern, corresponding to an electron dose of $1.2 \times 10^6$ $e^-$/Å$^2$. The camera length was set to 460 mm, and the probe was focused 20 nm above the specimen surface.

The twisted bilayer $SrTiO_3$ sample was investigated using a probe aberration-corrected JEOL NEOARM transmission electron microscope operated at 200 kV. The dataset was acquired using the Dectris Arina camera, yielding a four-dimensional STEM dataset with dimensions of 192 × 192 × 1600 × 1600. The scan step size was 0.50 Å, and the probe-forming semi-angle was 27.6 mrad. The beam current was 80 pA, and the dwell time was 34 μs per diffraction pattern, resulting in an estimated electron dose of $6.7 \times 10^4$ $e^-$/Å$^2$. The camera length was set to 12 cm, and the beam was focused ~20 nm above the sample surface.

### Data analysis and simulations

The convolution of a step function ($y = 0$ while $x < a$ and $y = 1$ while $x \geq a$) and Gaussian function $y = A\exp\left[-\frac{(x-\mu)^2}{2\sigma^2}\right] + B$ yields error function $\mathrm{erf}(x) = \frac{2}{\sqrt{\pi}}\int_0^x \exp(-t^2)\,dt$. A single error function, $y = A\,\mathrm{erf}\left(\frac{x-a}{\sqrt{2}\sigma}\right) + B$ is used to fit the edge of the depth-dependent phase profile of a single surface, while a superposition of two error functions, $y = A\left[\mathrm{erf}\left(\frac{x-a}{\sqrt{2}\sigma}\right) - \mathrm{erf}\left(\frac{x-b}{\sqrt{2}\sigma}\right)\right] + B$ is used for the whole two-surface sample profile, $b$ is the lower surface position and $a$ is the upper. Meanwhile, if the model is sufficiently accurate, the fitted depth aligns with the region where the slope reaches its maximum. In this way, we also reinforced the reliability of the fitting. Notably, the reconstruction performance was superior in the central regions compared to the edges, a discrepancy attributed to inherent algorithmic characteristics(*47*). To mitigate scan boundary-related uncertainties, we excluded certain boundary points during the fitting process used to determine the upper and lower surfaces. Using multivariate nonlinear regression fitting, depth-dependent phase profiles achieve comparable error (Fig. S11, Table1). Importantly, dilute point defects within the films only introduce localized phase disturbances without altering the surface positions. Therefore, the accuracy of the fitting can be ensured, since the perturbations in the intermediate phase are not of concern. It is worth noting that the sufficient number of iterations (usually thousands of iterations) is needed to ensure good convergence, especially for the depth-dependent phase magnitude during MEP reconstructions.

Simulations were performed under the frozen-phonon approximation with 100 phonon configurations using abTEM(*48*). We chose a model structure of $SrTiO_3$ with 20 TiO and 19 Sr atoms along each Ti or Sr column, which means a surface termination of $TiO_2$. The beam energy was 300 kV, step size was 0.37 Å, convergent semi-angle was 25 mrad, the data size was 256 × 256 × 100 × 100. The total electron dose was $1\times10^6$ e/Å$^2$, modeled with Poisson noise. The probe was overfocused by 20 nm above the sample. The reciprocal space sampling was 0.02561 Å$^{-1}$ per pixel.

**Acknowledgments**

**Funding:** This work was supported from National Key Research and Development Program of China (MOST, Grant Nos. 2023YFA1406400), the National Natural Science Foundation of China (Grant Nos. 52273227 and U22A6005), and Guangdong Major Project of Basic Research, China (Grant No. 2021B0301030003). J.-C.Y. acknowledges the financial support from the National Science and Technology Council (NSTC), Taiwan, under grant nos. 112-2112-M-006-020-MY3 and 114-2124-M-006-003.

**Author contributions:**

Conceptualization: Zhen Chen

Methodology: Zhen Chen, Huaicheng Yuan, Zehao Dong

Film growth: Yu-Chen Liu, Li-Shu Wang, and Jan-Chi Yang

Investigation: Huaicheng Yuan

Visualization: Huaicheng Yuan

Writing—original draft: Huaicheng Yuan

Writing—review & editing: Huaicheng Yuan, Zhen Chen

**Competing interests:**

Authors declare that they have no competing interests.

**Data and materials availability:**

All data are available in the main text or the supplementary materials.

## Figures and Tables

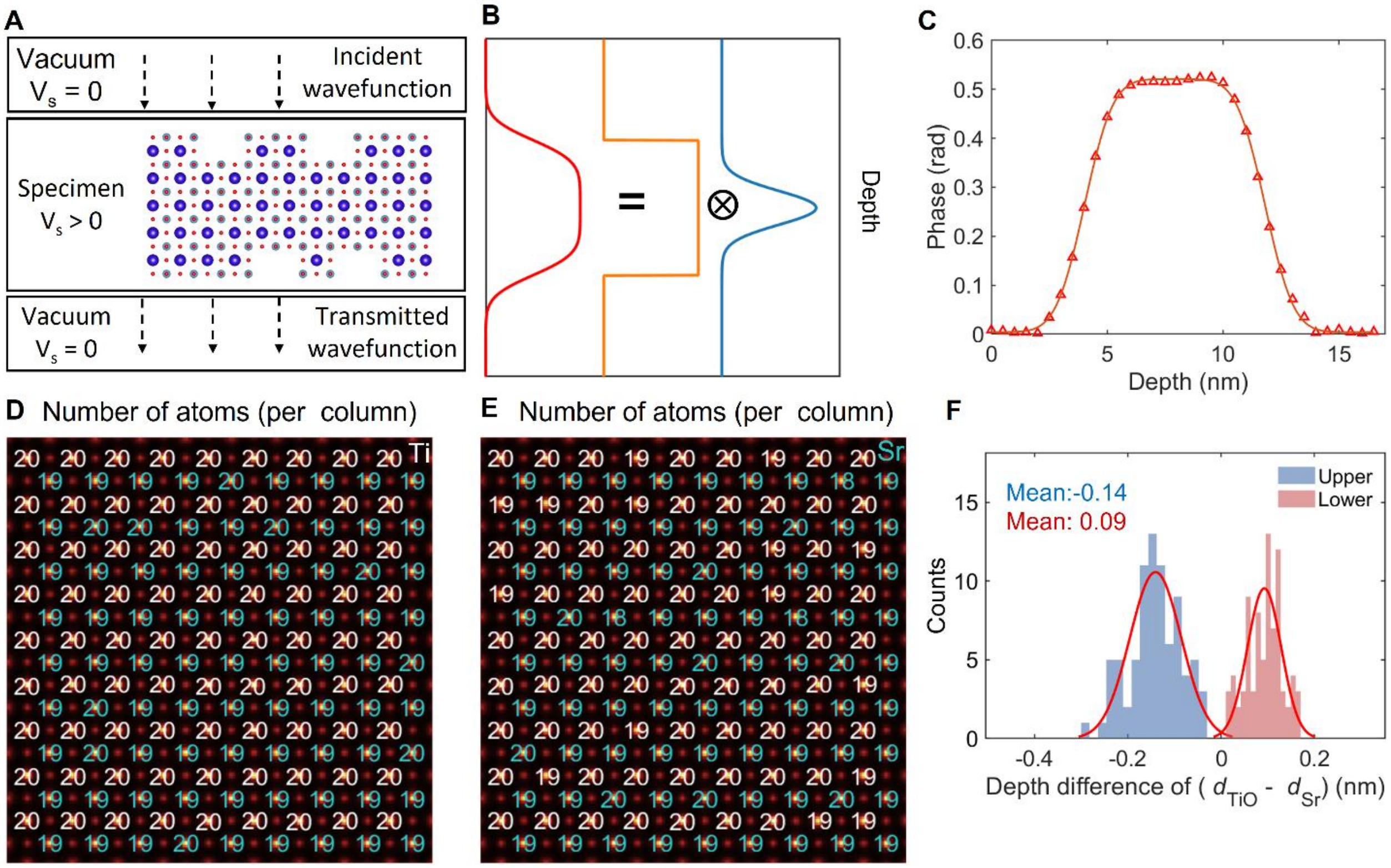

**Fig. 1. Principle of the surface morphology determination.** (**A**) Schematic illustration of the electron scattering. $V_s$ denotes the electrostatic potential. The electrostatic potential changes the phase of incident wave, then it can be reconstructed via multislice electron ptychography. The same surface termination of both upper and lower surfaces is adopted in the diagram (small light blue and red circles). (**B**) Phase depth profiles. From right to left are Gaussian function, step function and their convolution leading to the error function. (**C**) One depth profile of the phase and fitting function at TiO column from the simulation. (**D-E**) The number of TiO and Sr atoms at each atomic column. (**F**) Histogram of the difference between the depth of TiO atoms and the average number of their nearest-neighbor Sr atoms. Statistically, at the upper surface, $d_{\mathrm{TiO}} - d_{\mathrm{Sr}} < 0$ and at the lower surface, $d_{\mathrm{TiO}} - d_{\mathrm{Sr}} > 0$, which implies the $TiO_2$ termination.

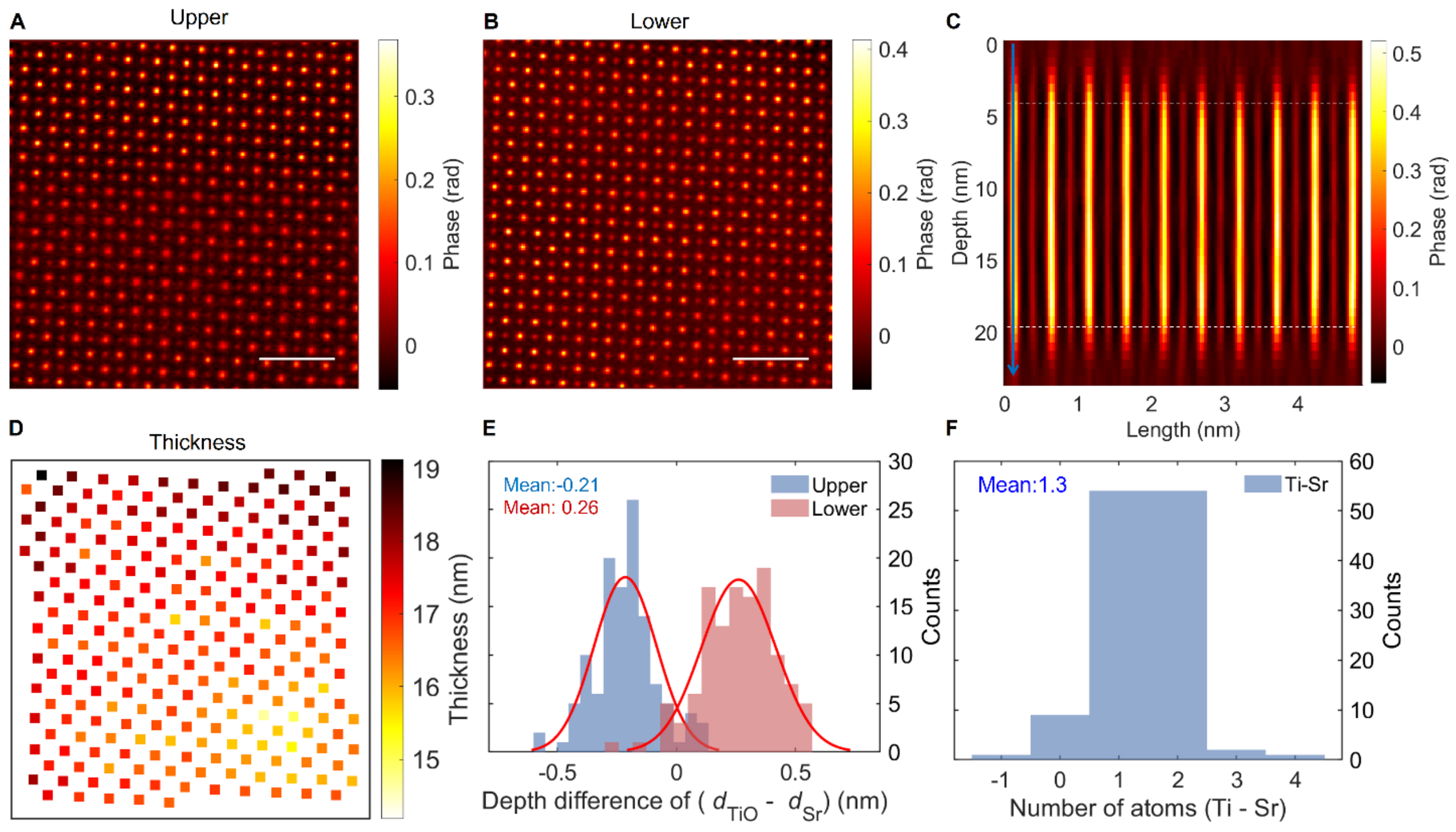

**Fig. 2. The surface topography of single crystalline $SrTiO_3$.** (**A-B**) Upper and lower slices of ptychographically reconstructed phase. (**C**) Cross section of phase distribution along the depth dimension. Dashed lines mark the rough depth positions of upper and lower surfaces, respectively. Blue solid line indicates the atomic column for a fitting of depth profile. (**D**) The thickness mapping of $SrTiO_3$, each pixel block represents a single atomic column. (**E**) Histogram of the difference between the depth of TiO atoms and the average number of their nearest-neighbor Sr atoms. Statistically, at the upper surface, $d_{\mathrm{TiO}} - d_{\mathrm{Sr}} < 0$ and at the lower surface, $d_{\mathrm{TiO}} - d_{\mathrm{Sr}} > 0$. (**F**) Histogram of the difference between the number of TiO atoms and the average number of their nearest-neighbor Sr atoms. Statistically, the number of TiO subtracts the number of Sr equals about 1, which implies the $TiO_2$ termination. Scale bar, 1 nm.

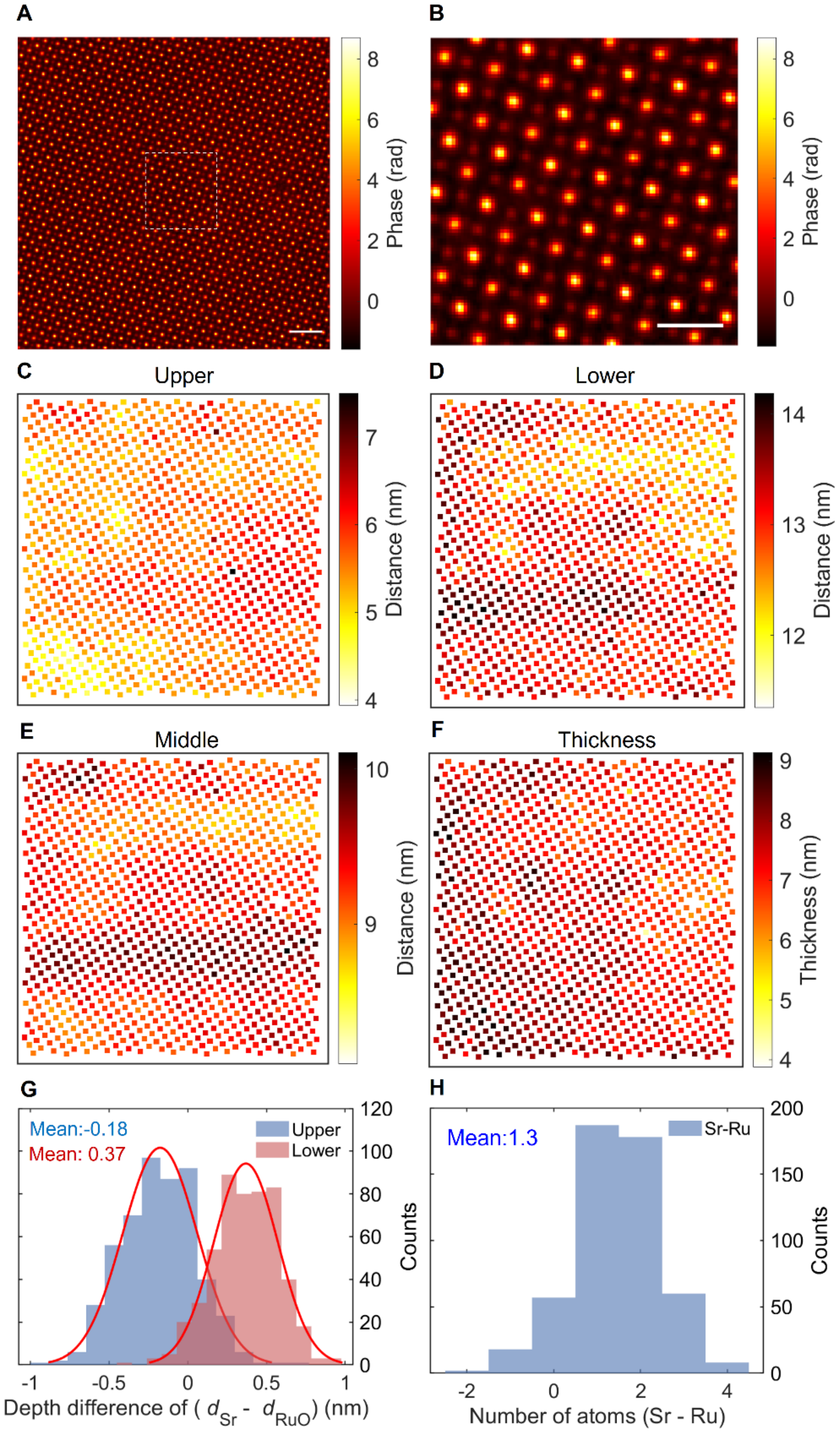

**Fig. 3. The surface topography of freestanding $SrRuO_3$.** (**A**) Summed phase of all slices. Scale bar, 1 nm. (**B**) Phase from the cropped region marked in (A). Scale bar, 0.5 nm. (**C-D**) The upper and lower surface topography of $SrRuO_3$. (**E**) The average depth position relative to reference plane (around the upper surface). (**F**) The thickness mapping of $SrRuO_3$. (**G**) Histogram of the difference between the depth of Sr atoms and the average depth of their four nearest-neighbor Ru atoms. Statistically, at the upper surface, $d_{Sr} - d_{RuO} < 0$ and at the lower surface, $d_{Sr} - d_{RuO} > 0$. (**H**) Histogram of the difference between the number of Sr atoms and the average number of their four nearest-neighbor RuO atoms. Statistically, the number of Sr minus the number of RuO equals about 1, which implies the relatively more SrO termination.

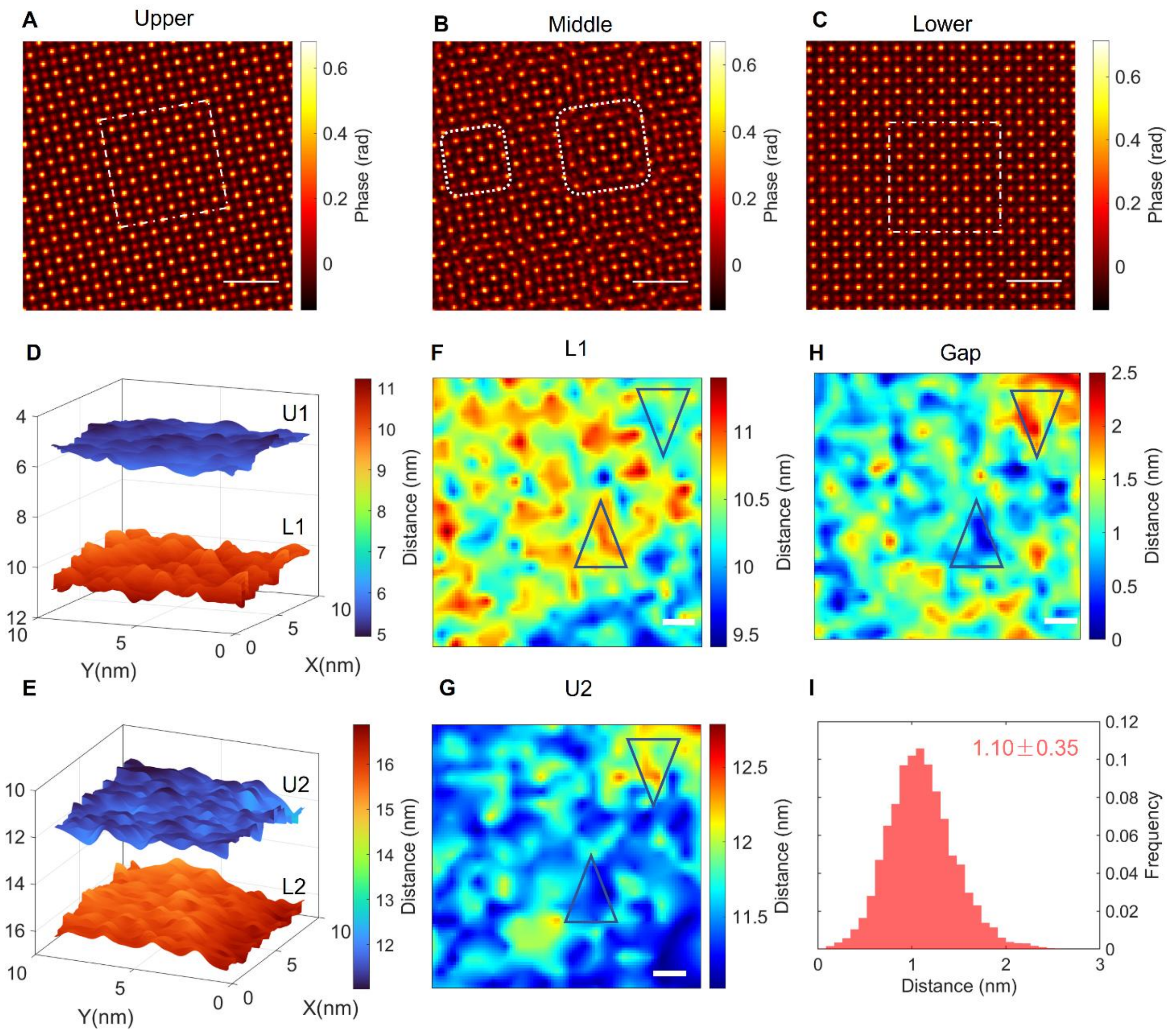

**Fig. 4. Topography of surfaces and interfaces of twisted bilayer $SrTiO_3$.** (**A-C**) Upper, middle, and lower slices of ptychographically reconstructed phase, respectively. The small dashed box in (B) marks a non-Moiré feature, while the large dashed box indicates the Moiré pattern. The twisted angle is 9.45°. (**D-E**) Surface topology of (A) and (C) of the twisted bilayer $SrTiO_3$ film from the same region. U1, upper 1, L1, lower 1; U2, upper 2, L2, lower 2. (**C-E**) Topography of L1, U2 and the gap between them. The triangles illustrate the interfacial inhomogeneity. (**I**) Histogram of the interfacial spacing. The average distance between two films is 1.10±0.35 nm. Scale bar, 1 nm.

# Supplementary Materials for

## Atomic-Scale Roughness of Freestanding Oxide Membranes Revealed by Electron Ptychography

Huaicheng Yuan *et al.*

Corresponding authors: zhen.chen@iphy.ac.cn

**The PDF file includes:**

Figs. S1 to S11.
Table S1.

**Other Supplementary Materials for this manuscript include the following:**

Movies S1 to S4.

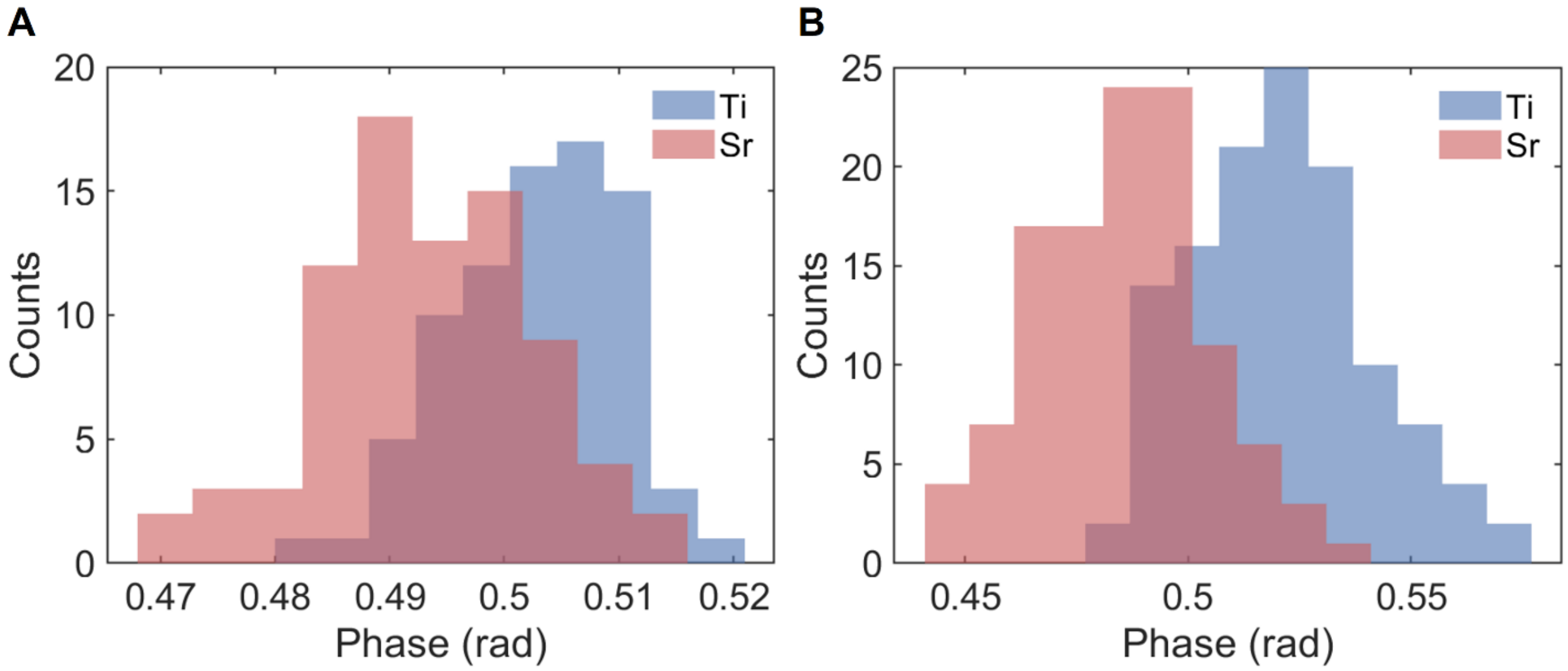

**Fig. S1. Phase statistics of $SrTiO_3$. (A)** The phase histogram of simulated $SrTiO_3$. The phase of TiO is 0.503±0.007 rad, slightly bigger than Sr, 0.494±0.009 rad. **(B)** The phase histogram of experimental $SrTiO_3$ single crystal. The phase of TiO is 0.521±0.020 rad, also bigger than Sr, 0.485±0.019 rad, indicating different kind of atoms.

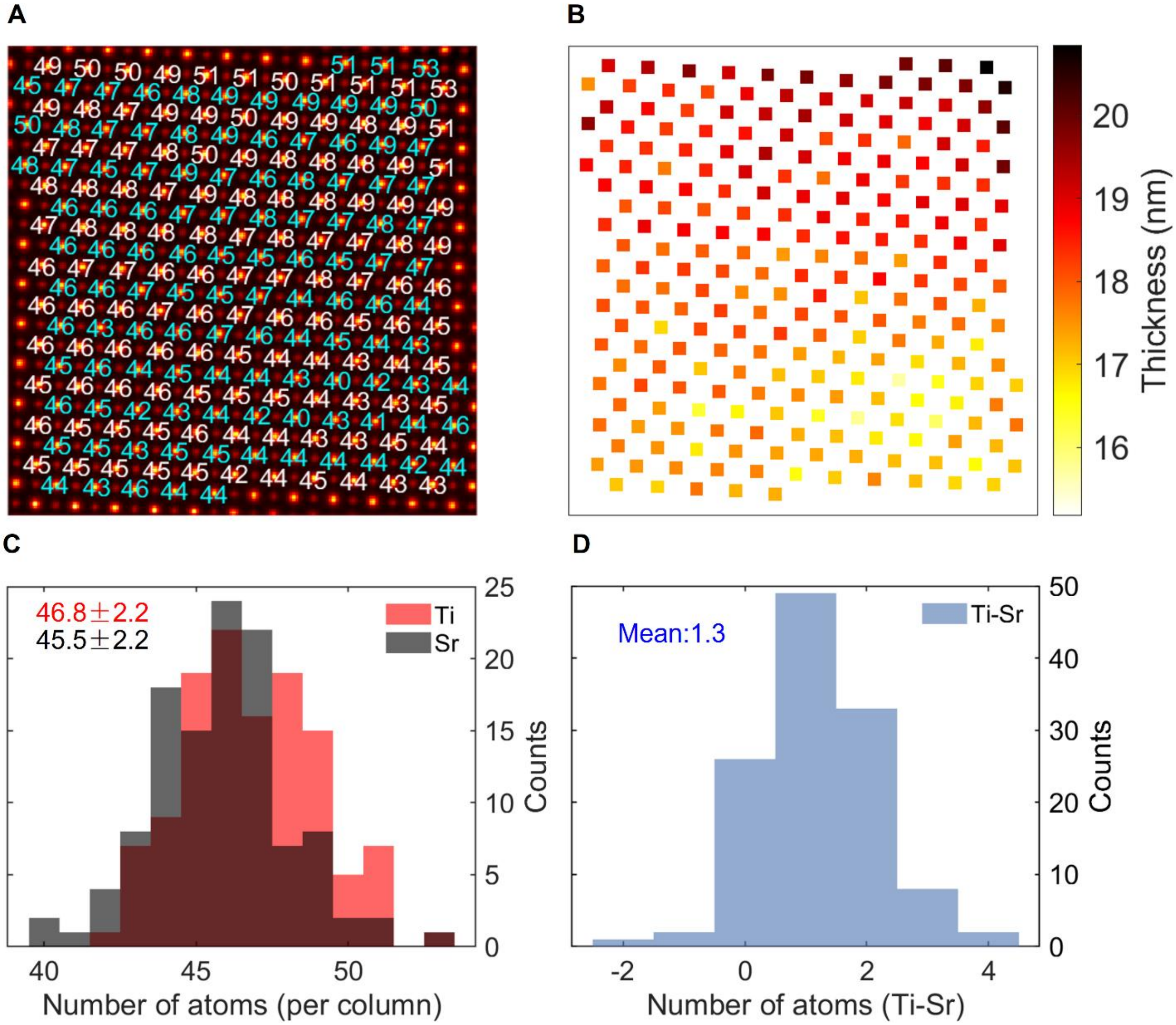

**Fig. S2. The thickness of $SrTiO_3$ determined from the summed phase. (A)** The number mapping of Sr and TiO atoms by summed phase analytical method. The colors of the numbers correspond to the same atoms as described in the main text of Fig.1. **(B)** The thickness mapping of corresponding atoms in (A) by summed phase analytical method. **(C)** The histogram of the number of atoms by summed phase analytical method. **(D)** Histogram of the difference between the number of TiO atoms and the average number of their nearest-neighbor Sr atoms. Statistically, the number of TiO minus the number of Sr equals about 1, which implies the $TiO_2$ termination.

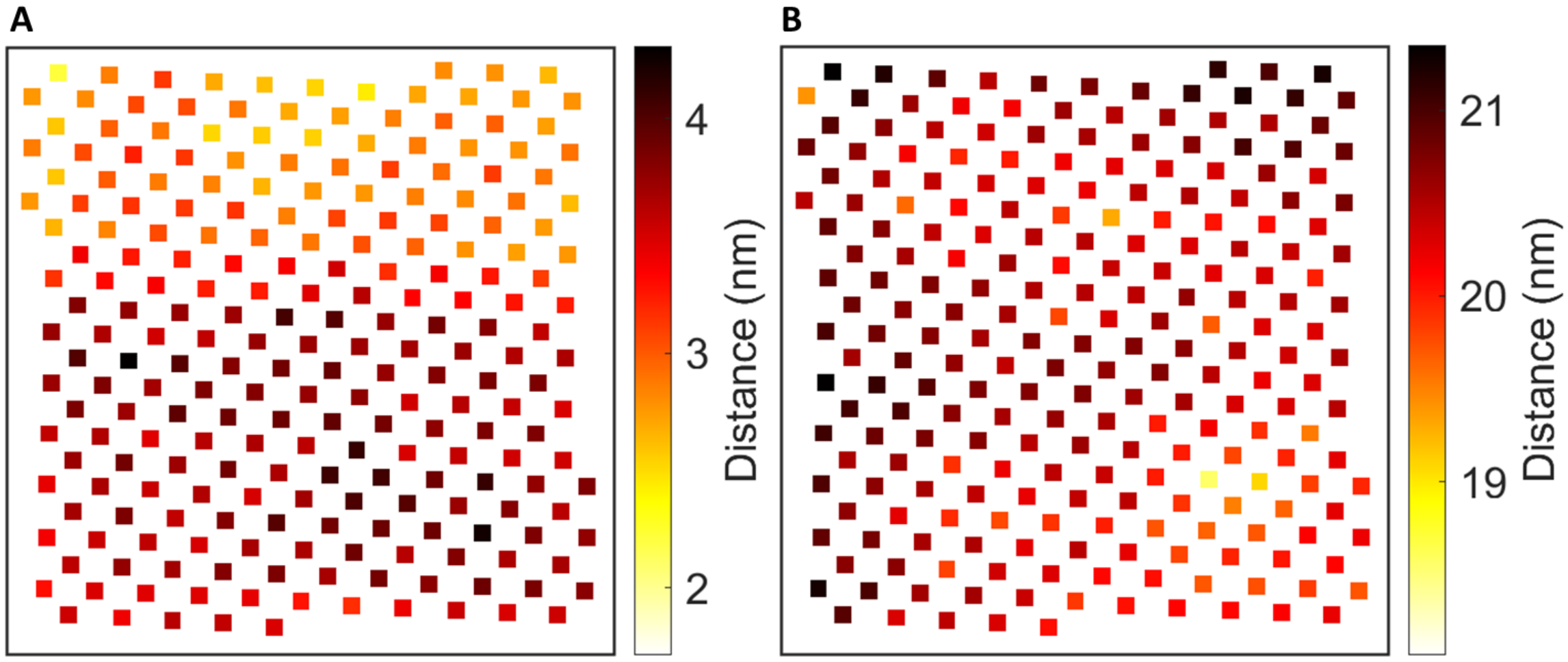

**Fig. S3. Upper and lower surface of single crystalline $SrTiO_3$ from experiments. (A)** Upper surface of $SrTiO_3$. The depth position is 3.28±0.43 nm. **(B)** Lower surface of $SrTiO_3$. The depth position is 20.55±0.35 nm. The edge-like surface came from FIB process.

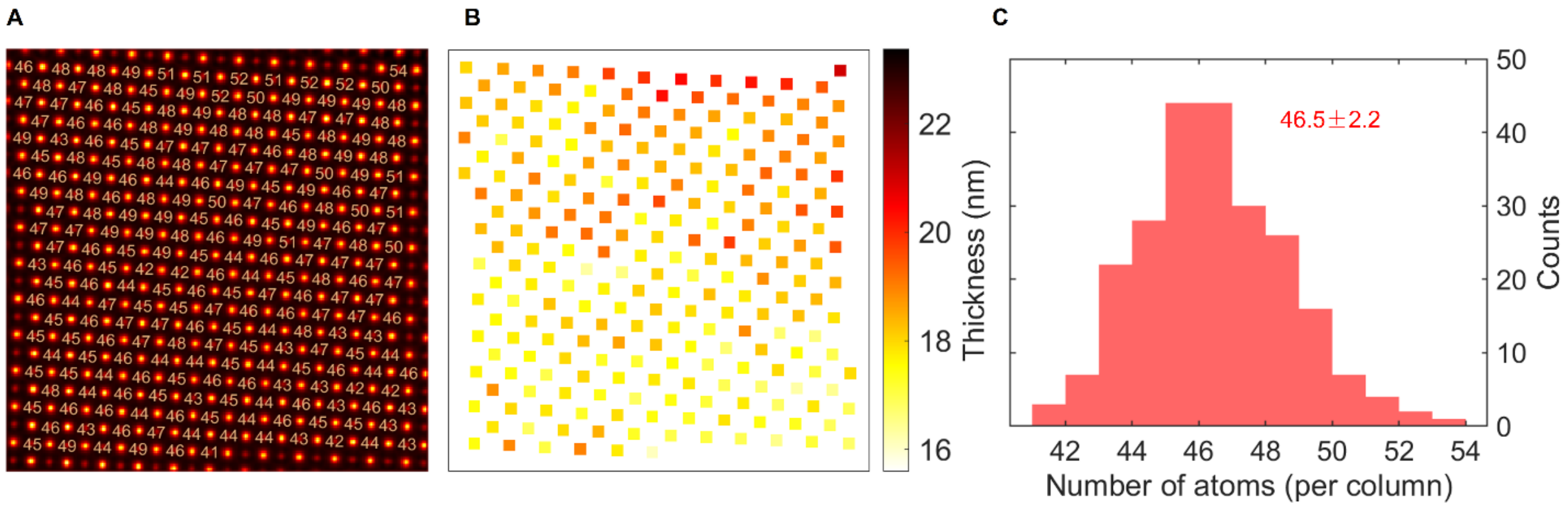

**Fig. S4. The number of O atoms determined from the summed phase. (A)** The mapping of O atom number calculating from summed phase divided by the phase from single slice. **(B)** The thickness mapping of O atoms calculating from the summed phase divided by the phase from single slice. **(C)** Histogram of the number of O atoms, its quantity is consistent with that of TiO in Fig. 2B, indicating the $TiO_2$ termination.

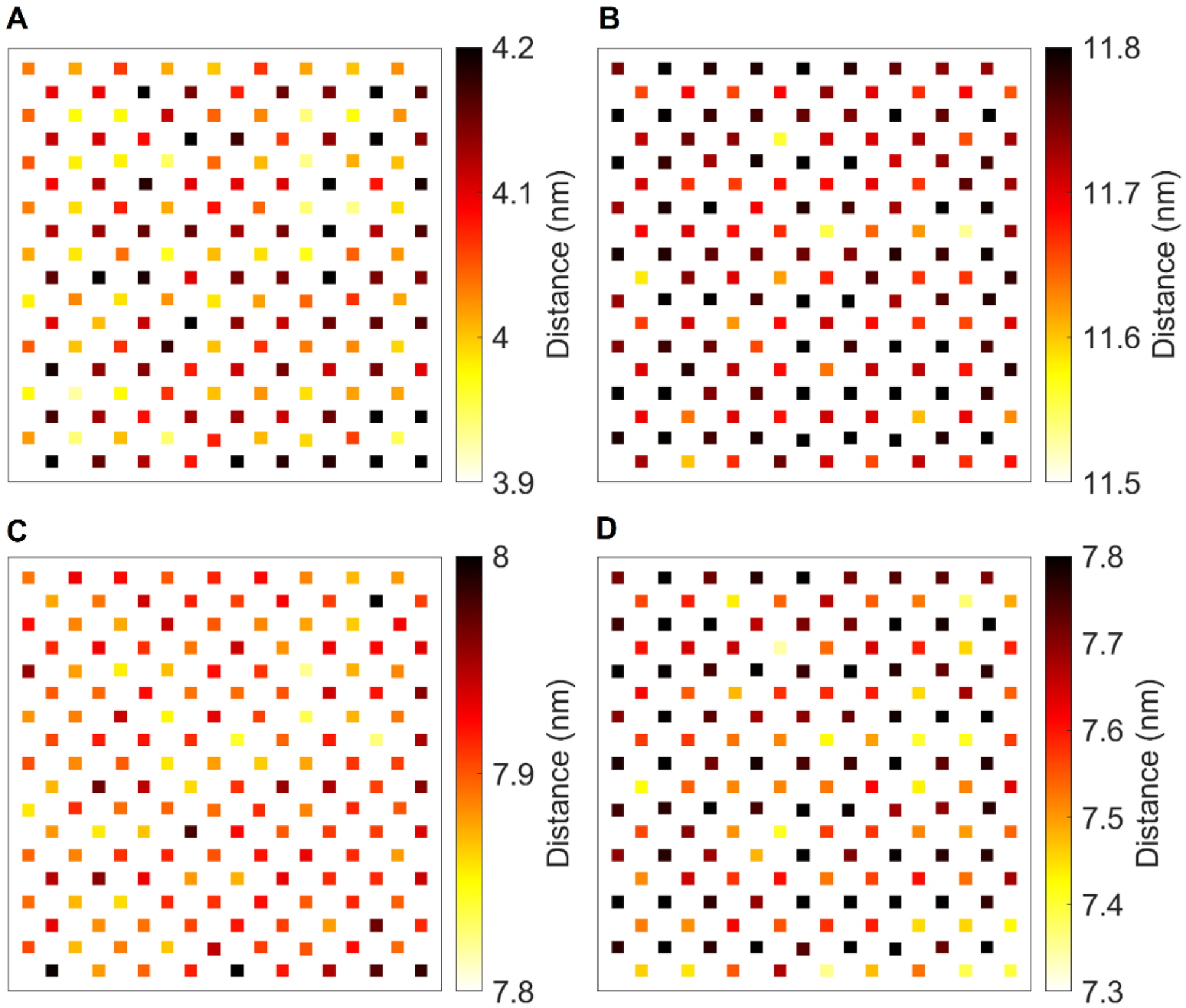

**Fig. S5. The surfaces of simulated $SrTiO_3$ by fitting method. (A)** Upper surface of $SrTiO_3$. The depth position is 4.01±0.04 nm. **(B)** Lower surface of $SrTiO_3$. The depth position is 11.78±0.03 nm. **(C)** The average depth position relative to reference plane. The average depth is 7.89±0.05 nm. **(D)** The thickness mapping of $SrTiO_3$. The thickness is 7.77±0.05 nm.

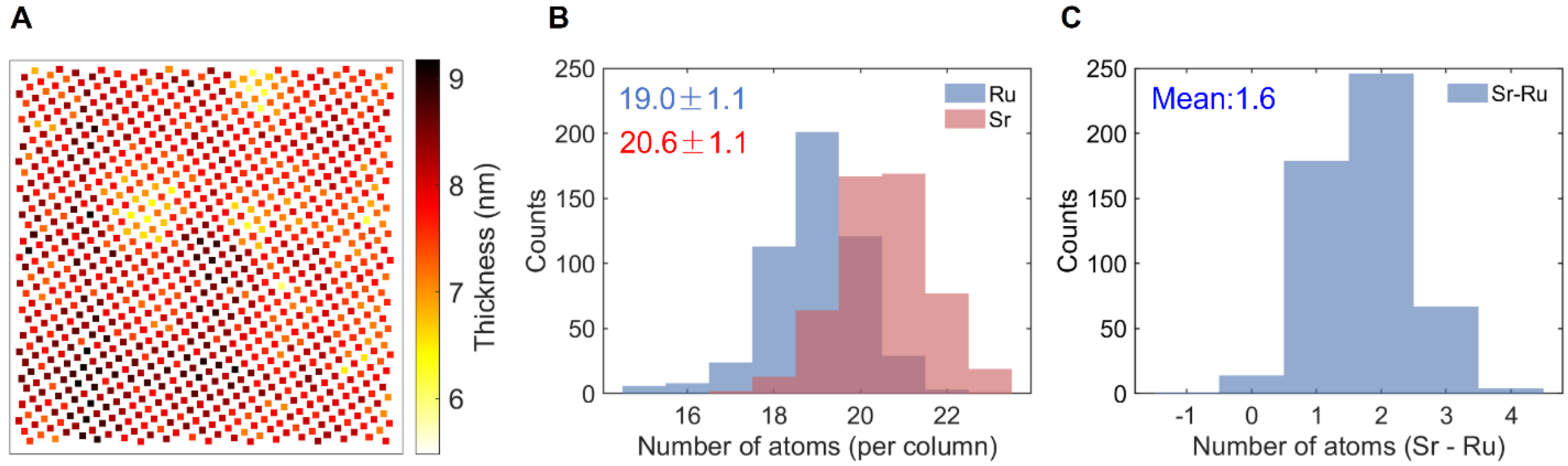

**Fig. S6. The thickness of freestanding $SrRuO_3$ by summed phase. (A)** The thickness mapping of atoms by summed phase analytical method. **(B)** The histogram of the number of atoms by summed phase analytical method. **(C)** Histogram of the difference between the number of Sr atoms and the average number of their nearest-neighbor RuO atoms. Statistically, the number of Sr minus the number of RuO equals about 1, which implies the SrO termination.

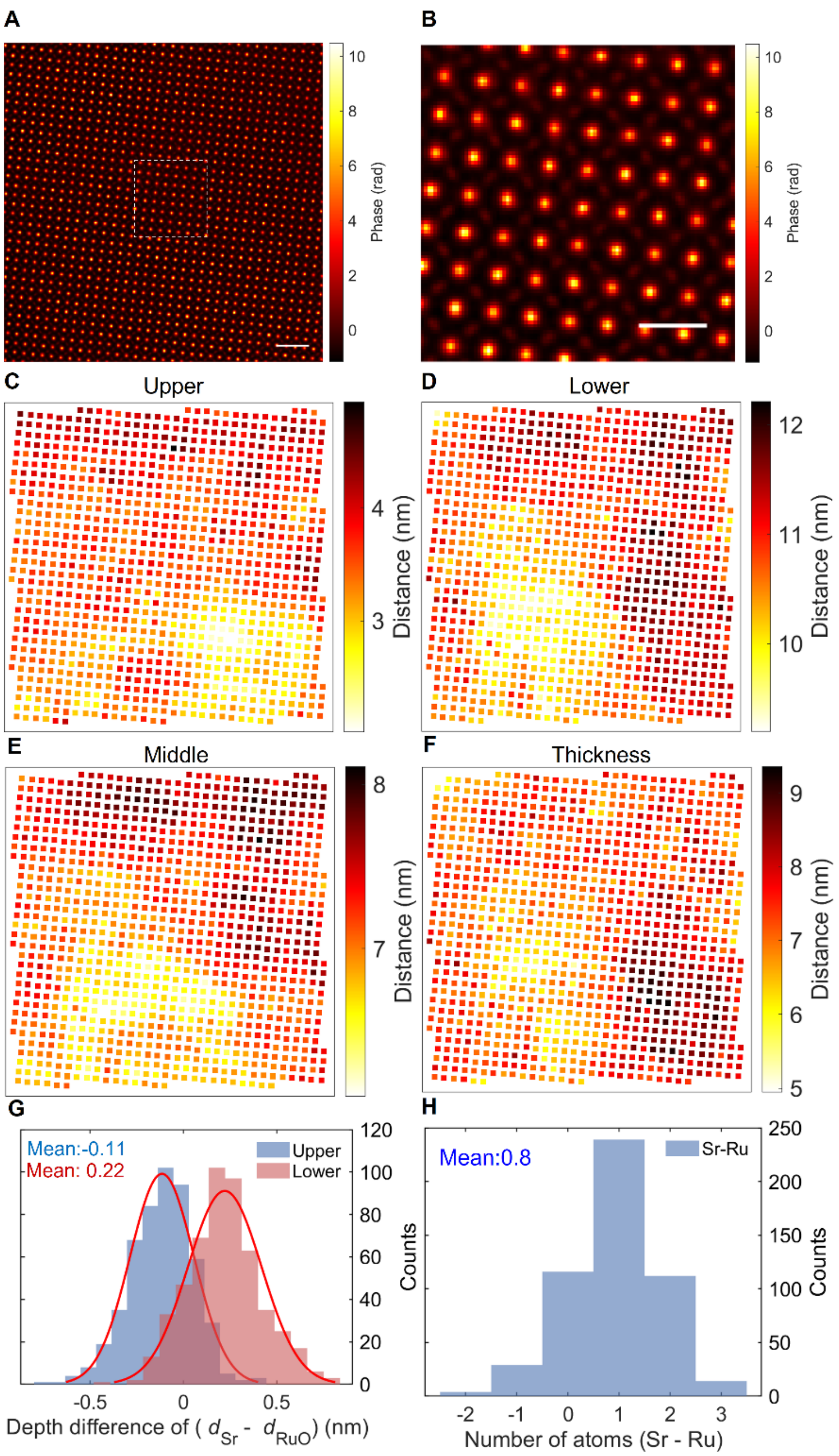

**Fig. S7. The surface topography of $SrRuO_3$ from another sample region B.** (**A**) Summed phase of all slices. Scale bar, 1 nm. (**B**) Phase from the cropped region marked in (A). Scale bar, 0.5 nm. (**C-D**) The upper and lower topography of $SrRuO_3$. (**E**) The average depth position relative to reference plane. (**F**) The thickness mapping of $SrRuO_3$. Localized bending was identified in regions where the local average depth position deviated from surrounding areas without significant thickness variation. (**G**) Histogram of the difference between the depth of Sr atoms and the average number of their nearest-neighbor RuO atoms. Statistically, at the upper surface, $d_{Sr} - d_{RuO} < 0$ and at the lower surface, $d_{Sr} - d_{RuO} > 0$. (**H**) Histogram of the difference between the number of Sr atoms and the average number of their nearest-neighbor RuO atoms. Statistically, the number of Sr minus the number of RuO equals about 1, which implies the relatively more SrO termination.

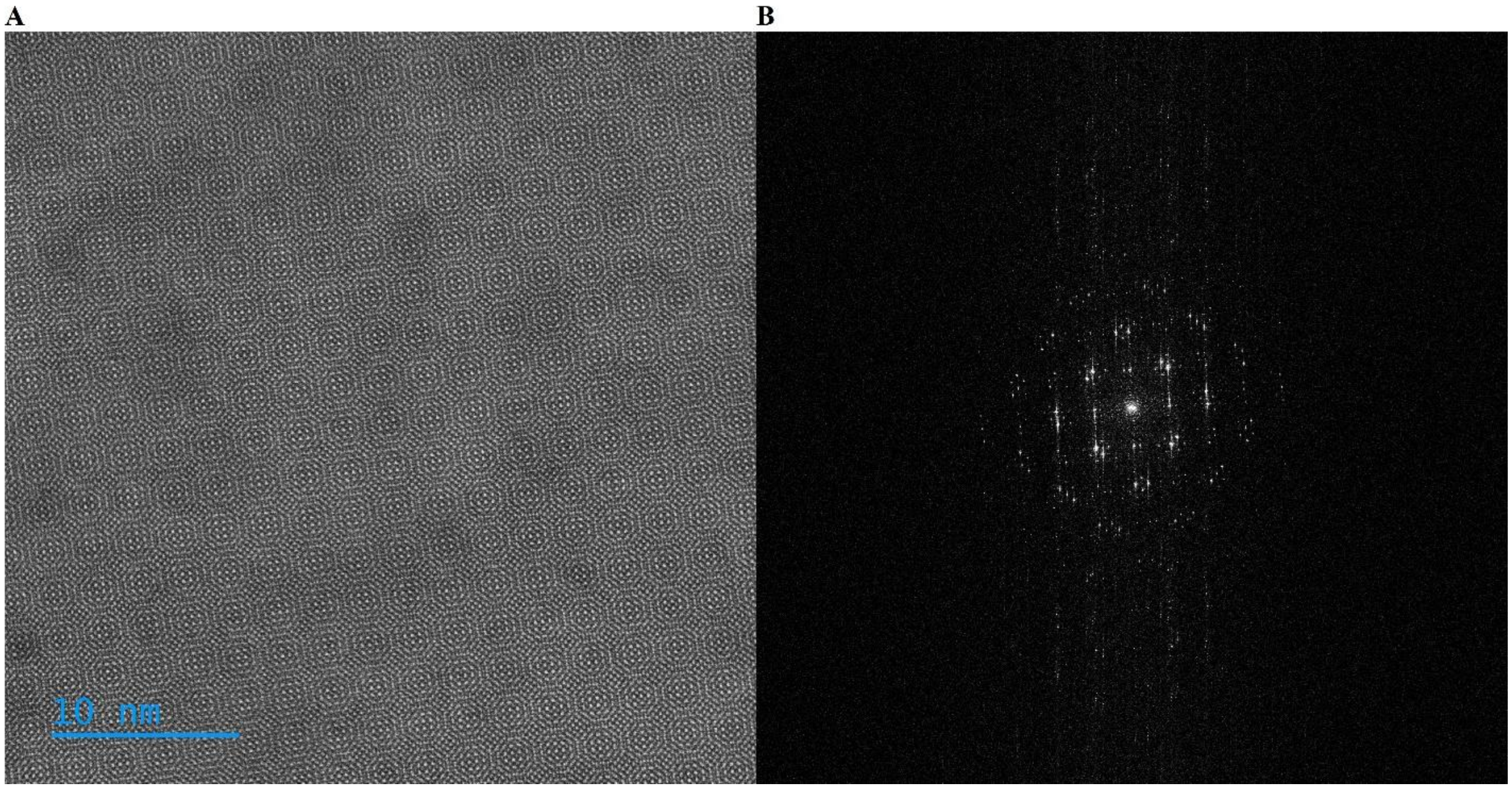

**Fig. S8. HAADF of twisted bilayer $SrTiO_3$ and FFT. (A)** HAADF of twisted bilayer $SrTiO_3$. It is difficult to resolve the atomic structural details and the three-dimensional information is lacking. **(B)** FFT of (A). The two sets of FFT spots indicate the coexistence of two distinct structures.

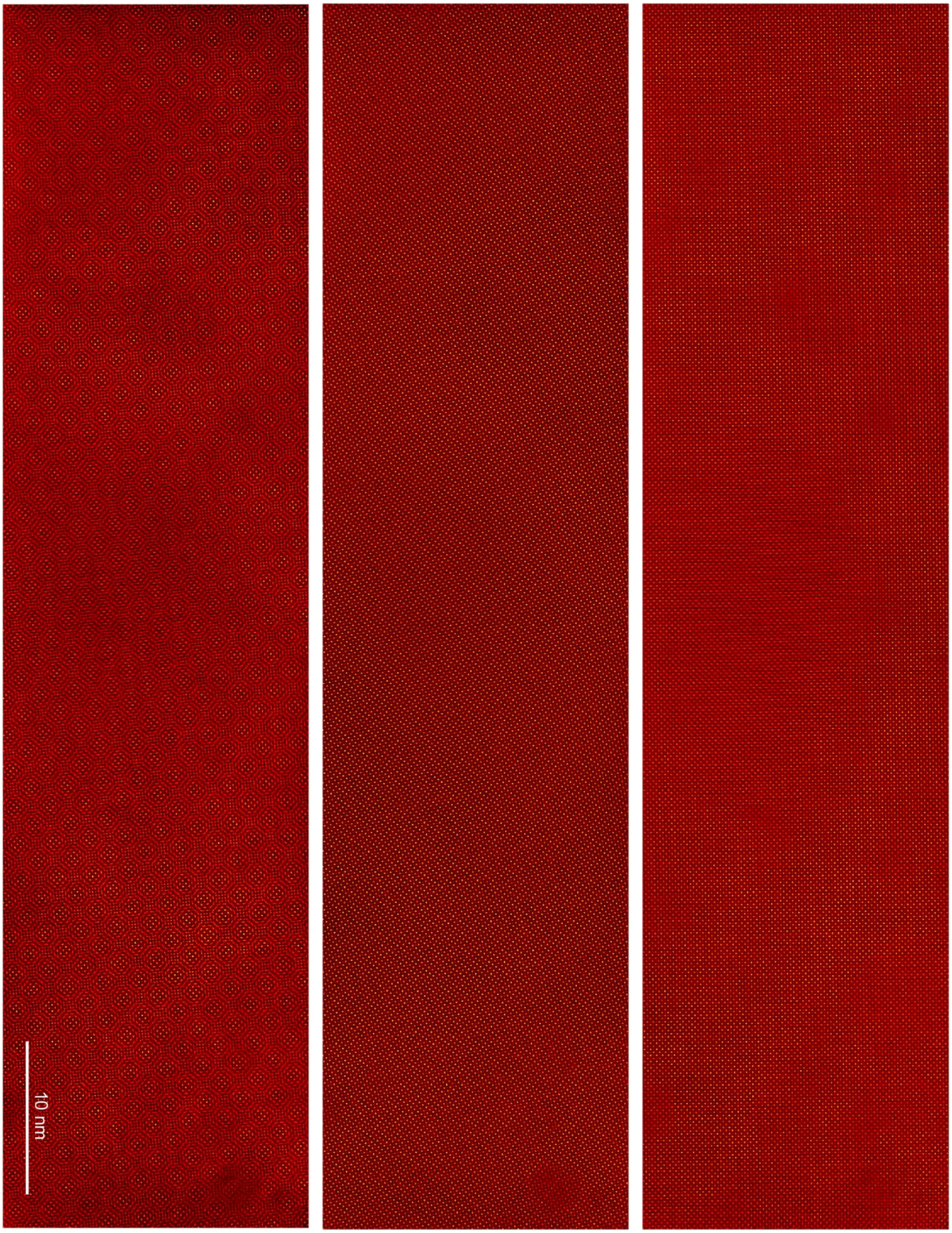

**Fig. S9. Multislice electron ptychographic reconstruction of twisted bilayer $SrTiO_3$.** From left to right are sum, upper and lower phase, respectively. The non-uniform contrast arises from variations in thickness and crystal orientation.

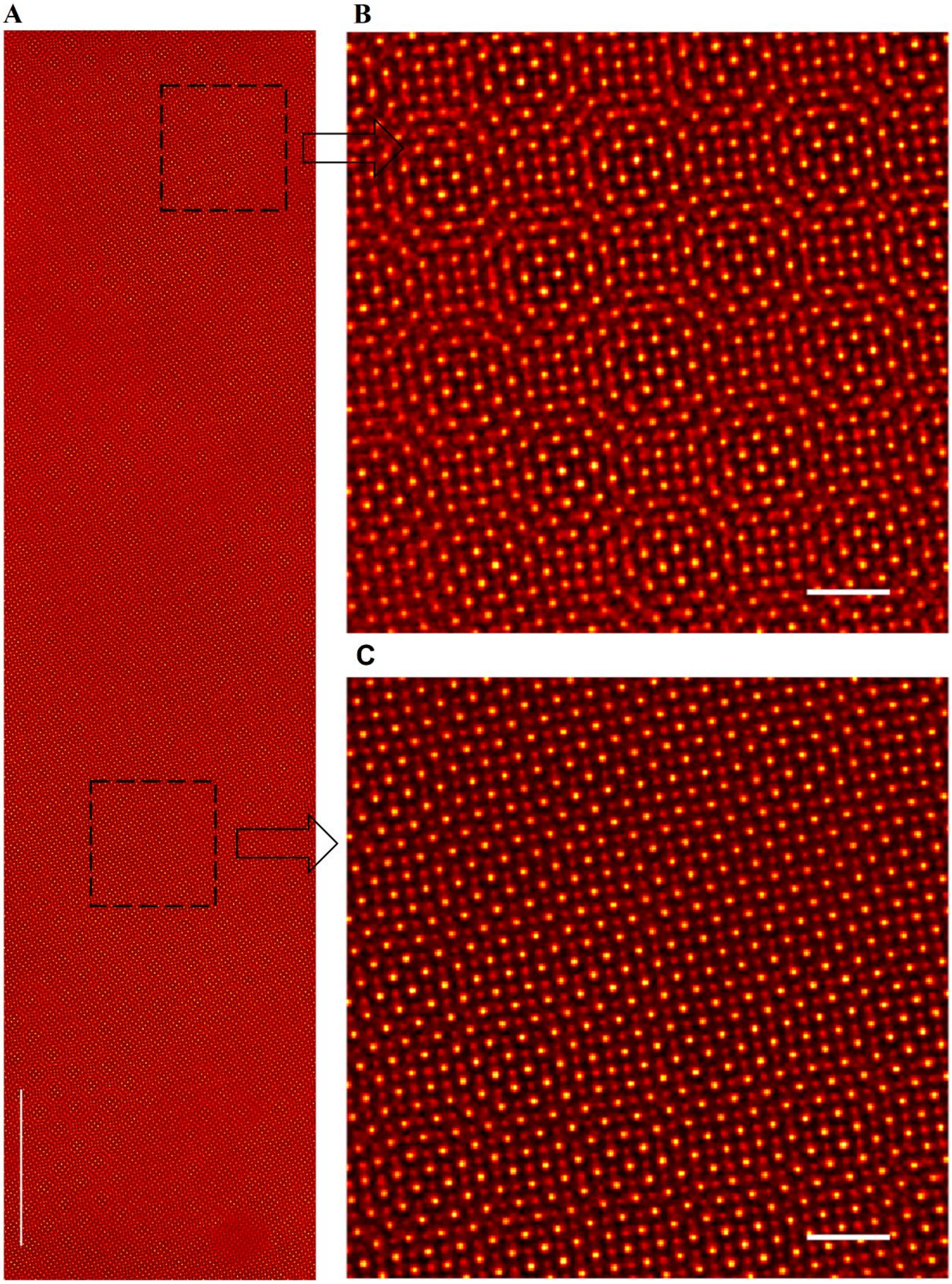

**Fig. S10. Multislice electron ptychographic reconstruction of twisted bilayer $SrTiO_3$. (A)** Middle phase of the twisted bilayer $SrTiO_3$. Scale bar, 10 nm. **(B-C)** Phase from the cropped region marked in (A). The structural differences between (B) and (C) originate from surface roughness. Scale bar, 1 nm.

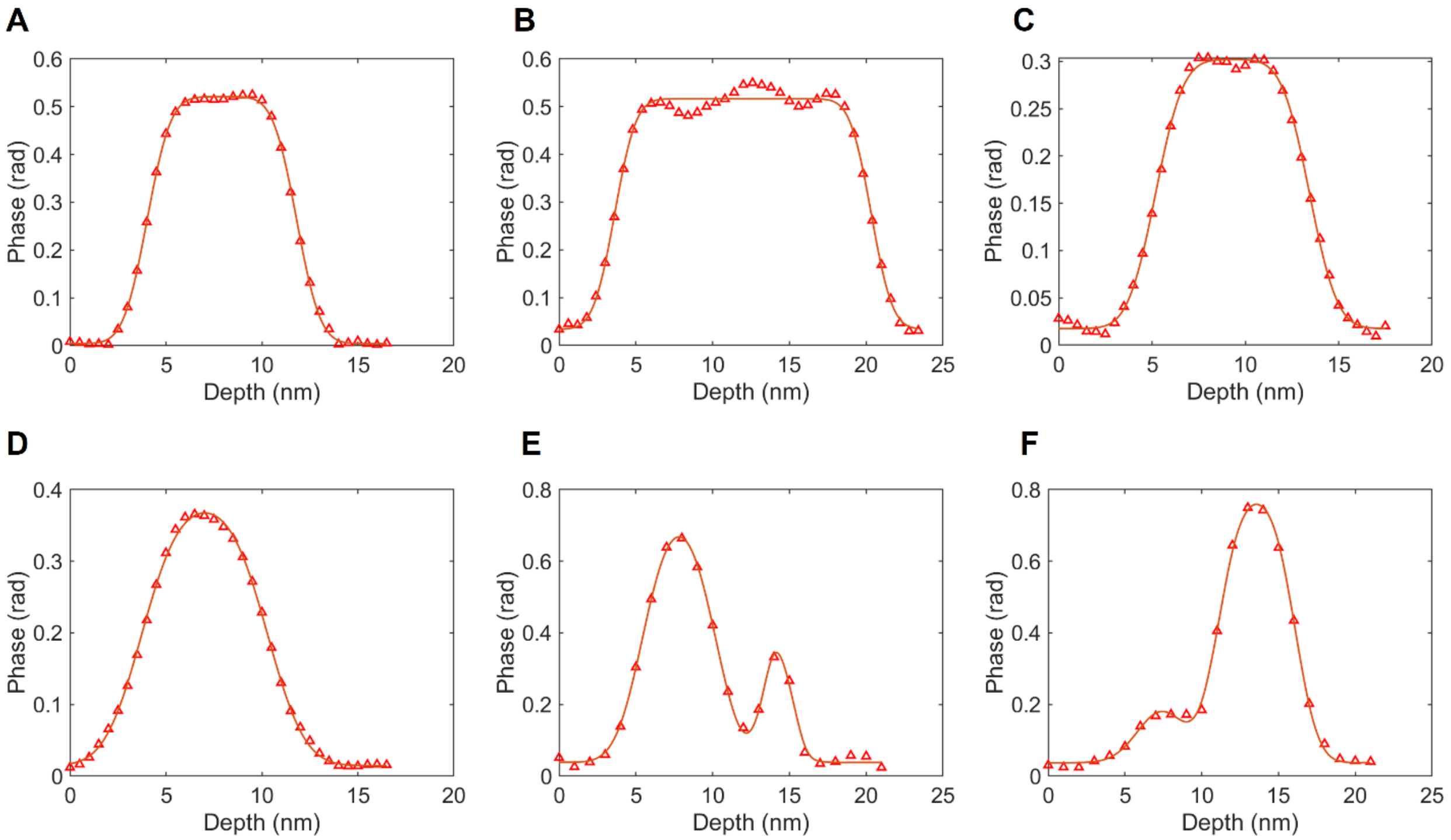

**Fig. S11. Different types of phase depth profiles for fitting. (A)** Phase to depth profile of simulated $SrTiO_3$ of Ti column with 0.2 Å fitting error of depth position. **(B)** Phase to depth profile of experimental $SrTiO_3$ of Ti column with 0.8 Å fitting error of depth position. **(C)** Phase to depth profile of $SrRuO_3$ region A of Sr columns with 0.6 Å fitting error of depth position. **(D)** Phase to depth profile of $SrRuO_3$ region B of Sr column with 0.7 Å fitting error of depth position. **(E-F)** Phase to depth profile of twisted bilayer $SrTiO_3$ TiO columns with 0.8 Å fitting error of depth position, (E) from first layer and (F) from second layer. The fitting function for the other heavy element is similar with comparable error.

**Table. S1 Depth position of sample surfaces and fitting errors**

| | Upper surface (nm) | Lower surface (nm) | Thick (nm) | Average (nm) | Fitting error (nm) |
|---|---|---|---|---|---|
| Simulated $SrTiO_3$ | 4.01±0.04 | 11.78±0.03 | 7.77±0.05 | 7.89±0.05 | 0.02 |
| Experimental $SrTiO_3$ | 3.28±0.43 | 20.55±0.35 | 17.27±0.63 | 11.92±0.24 | 0.08 |
| $SrRuO_3$ A | 5.28±0.41 | 12.92±0.40 | 7.64±0.62 | 9.10±0.26 | 0.07 |
| $SrRuO_3$ B | 3.43±0.44 | 10.93±0.56 | 7.50±0.66 | 7.18±0.38 | 0.06 |
| Twisted bilayer $SrTiO_3$ first layer | 5.44±0.22 | 10.45±0.38 | 5.01±0.41 | 7.95±0.23 | 0.08 |
| Twisted bilayer $SrTiO_3$ second layer | 11.58±0.33 | 15.95±0.36 | 4.37±0.42 | 13.77±0.27 | 0.08 |